# Studies and analysis of reference management software: a literature review

Jesús Tramullas
Ana Sánchez-Casabón
{jesus,asanchez}@unizar.es
Dept .of Library & Information Science, University of Zaragoza
Piedad Garrido-Picazo
piedad@unizar.es
Dept. of Computer and Software Engineering, University of Zaragoza

**Abstract**: Reference management software is a well-known tool for scientific research work. Since the 1980s, it has been the subject of reviews and evaluations in library and information science literature. This paper presents a systematic review of published studies that evaluate reference management software with a comparative approach. The objective is to identify the types, models, and evaluation criteria that authors have adopted, in order to determine whether the methods used provide adequate methodological rigor and useful contributions to the field of study.

## 1. Introduction and background

Reference management software has been a useful tool for researchers since the 1980s. In those early years, tools were made ad-hoc, and some were based on the dBase II/III database management system (Bertrand and Bader, 1980; Kunin, 1985). In a short period of time a market was created and commercial products were developed to provide support to this type of information resources. The need of researchers to systematize scientific literature in both group and personal contexts, and to integrate mechanisms into scientific production environments in order to facilitate and expedite the process of writing and publishing research results, requires that these types of applications receive almost constant attention in specialized library and information science literature.

The result of this interest is reflected, in bibliographical terms, in the publication of numerous articles almost exclusively devoted to describing, analyzing, and comparing the characteristics of several reference management software products (Norman, 2010). Using these studies it is possible to trace use and consumption patterns in scientific information along with production and publication processes. The review and study of the contents of these articles, over an extended period of time, can provide data on the evolution of user needs, industrial life cycles, and the development of evaluation processes of software tools geared toward scientific information management.

A significant number of publications have focused on providing researchers and professionals with the information needed to decide which tool is best to meet their needs. To that end, several authors have provided individual or overall assessments of the tools available, in accordance with several criteria and approaches. Although Moore (1991) originally suggested an evaluation grid for highlighting features in reference management software, it is only recently that some basic proposals have been published in order to develop specifically designed evaluations for these tools (Marino, 2012).

## 2. Goals and working hypotheses

As mentioned earlier, the main goal of this article is to identify and analyze the guidelines, processes, and evaluation techniques used in articles when evaluating reference management software. A secondary goal of this article is to identify the scientific area of publication where the various articles reviewing reference management software are located. Lastly, this research will provide an overview of the temporal evolution of functions provided by reference management software tools.

The working hypothesis for this article is that published assessments of reference management software are not based on rigorous methodological analysis, which ultimately degrades the quality of the publications and the value such articles might have for their intended audiences.

## 3. Methodology

The method adopted for this article follows the traditional approach of a systematic literature review. To that end, the eight steps formulated and suggested by Okoli and Schabram (2010) for systematic literature reviews of formulated information systems were followed. The research goals were defined and the information and required data were collected and subsequently tabulated, ordered, reviewed, and analyzed. This process allows articles to be characterized in accordance with several domains. Finally, the findings were compared with the established working hypothesis.

## 4. Data collection

A literature search of the databases *Web Of Science* and *Scopus* was conducted in March 20, 2015, using the search expressions TITLE-ABS-KEY("citation management software") OR TITLE-ABS-KEY("reference management software") OR TITLE-ABS-KEY("citation tools") OR TITLE-ABS-KEY("citation reference"). *Web of Science* produced 58 results, 32 of which were selected. *Scopus* produced 130 results, the subsequent review of which narrowed down the selection to 55. Once the core of articles suitable for evaluation was identified, the process was extended to identify other publications through *Google Scholar* and *Dialnet*. The search for references was completed using the bibliographies of suitable articles, by searching specialized interest groups in *Mendeley* and *Zotero*, creating a final list of 100 possible articles.

Once the references were organized and refined, a systematic review process was followed in order to identify those articles fitting the parameters of this study. To that end, the selected articles were those providing a review or description of two or more reference management software tools; articles that focused on news, tutorials, or specific tool descriptions were rejected. The process identified 37 articles that provided a description or comparison between two or more reference management software tools. For each particular article, information regarding authorship, year; and journal; conference proceedings; area of knowledge of the journal; and type and evaluation characteristics was provided.

## 5. Results analysis and discussion

The 37 articles reviewed (table 1) were published between 1987 and 2014. There is not a particular year in which more studies were published, although publication frequency increased for a decade beginning in 2000. Four studies were identified between 2011 and 2014.

| Author(s) | Year | Journal | Knowledge area | Type of review |
|---|---|---|---|---|
| Gurney, Wigton | 1987 | *American Journal of Roentgenology* | Biomedicine | Description |
| Brantz, Galla | 1988 | *Bull Med Libr Assoc* | Biomedicine | Comparative description |
| Garfield, Flanagan, Fox | 1989 | *Journal of Clinical Monitoring and Computing* | Biomedicine | Quantitative comparison |
| Nashelsky, Earley | 1991 | *Library Software Review* | Library&Inform. Science | Description |
| Jones | 1993 | *BMJ* | Biomedicine | Description |
| Miller | 1994 | *MD Computing: computers in medical practice* | Biomedicine | Comparative description |
| Cibbarelli | 1995 | *Computers in Libraries* | Library&Inform. Science | Comparative description |
| Tramullas | 1995 | *Tendencias de Investigación en Documentación.* | Library&Inform. Science | Comparative description |
| Nico,, Ouellette, Bird, Harper, Kelley | 1996 | *Computers in Nursing* | Biomedicine | Comparative description |
| Bravo | 1996 | *El Profesional de la Información* | Library&Inform. Science | Description |
| Bravo, Astorga | 2000 | *Atención Primaria* | Biomedicine | Description |
| Shapland | 2000 | *Library and Information Briefings* | Library&Inform. Science | Comparative description |
| Koopman | 2002 | *Internet Reference Services Quarterly* | Library&Inform. Science | Comparative description |
| May | 2003 | *The Scientist* | Multidisciplinar | Description |
| Kessler, Van Ullen | 2005 | *The Journal of Academic Librarianship* | Library&Inform. Science | Quantitative comparison |
| Mattison | 2005 | *Searcher* | Library&Inform. Science | Comparative description |
| Duarte García | 2007 | *El Profesional de la Información* | Library&Inform. Science | Comparative description |

| Smith, Baker | 2007 | *International Journal of Mental Health Nursing* | Biomedicine | Comparative description |
| --- | --- | --- | --- | --- |
| Giménez, Tramullas | 2007 | *IX Jornadas Españolas de Documentación* | Library&Inform. Science | Comparative description |
| Hernández, El-Masri, Hernández | 2008 | *Diabetes Educator* | Biomedicine | Comparative description |
| Gomis, Gall, Brahmi | 2008 | *Medical Reference Services Quarterly* | Biomedicine | Comparative description |
| Cordón, Martín, Alonso | 2009 | *El Profesional de la Información* | Library&Inform. Science | Comparative description |
| Butros, Taylor | 2010 | *36th IAMSLIC Annual Conference* | Library&Inform. Science | Description |
| Fenner | 2010 | *Cellular Therapy and Transplantation* | Biomedicine | Description |
| Alonso | 2010 | *ThinkEPI* | Library&Inform. Science | Description |
| Gilmour, Cobus-Kuo | 2011 | *Issues in Science and Technology Librarianship* | Library&Inform. Science | Comparative description |
| Kern, Hensley | 2011 | *Reference & User Services Quarterly* | Library&Inform. Science | Description |
| Rapp | 2011 | *Library Journal* | Library&Inform. Science | Description |
| Web, Platter | 2011 | *UKOLUG Newsletter* | Library&Inform. Science | Comparative description |
| Glassman, Sorensen | 2012 | *Journal of Electronic Resources in Medical Libraries* | Biomedicine | Comparative description |
| Zhang | 2012 | *Medical Reference Services Quarterly* | Biomedicine | Comparative description |
| Mahajan, Hogarth | 2013 | *Chest* | Biomedicine | Comparative description |
| Steeleworthy, Dewan | 2013 | *Partnertship: the Canadian Journal of Library and Information Practice and Research* | Library&Inform. Science | Comparative description |
| Homol | 2014 | *The Journal of Academic Librarianship* | Library&Inform. Science | Comparative description |
| Casado, Maroto, Dani, Ávila | 2014 | *FMC - Formación Médica Continuada en Atención Primaria* | Biomedicine | Description |
| Yamakawa, Kubota, beuren, Scalvenzi, | 2014 | *Transinformacao* | Engineering | Description |
| Cauchick Basak | 2014 | *Journal of Economics & Behavioral Studies* | Engineering | Quantitative comparison |

Table 1. Reference management software revised papers.

From the 37 texts reviewed, only three of them were published in conference proceedings. The rest were published in specialized journals: none of the journals of publication are more significant with regard to the amount of articles published. There is only one exception, *El Profesional de la Información (EPI)*, a journal that published four of the articles reviewed; in addition, a fifth article was published in its related publication *ThinkEPI*. Regarding the language in which they were written, nine articles were in Spanish and published between 1995 and 2014.

Each journal and article was assigned to an area of knowledge. Applying the most accepted division, the areas of knowledge used were Biomedicine, Social Sciences and Law, Engineering and Computer Science, and Humanities. In addition, the specific area of Library and Information Science was created. Data show that most of the articles focus on the areas of Biomedicine (15 articles, 40.5%) and Library and Information Science (19 articles, 51.4%), which amount to 91.9% of all the articles. If the series is reviewed chronologically, there were slightly more articles published in the area of Library and Information Science from 1995 onward.. The high percentage of publications in Biomedicine and Library and Information Science indicates a particular concern or interest in both communities regarding the state, evolution, and use of software management tools.

The evaluations conducted by the authors of the 37 articles are the main object of analysis and discussion in this article. Two levels of review were used to study the articles:

1. The first level corresponds to the type of review conducted by the authors of each study. Three types were identified:

    - Description: articles where authors described various aspects of the software management tools, but did not include a systematic comparison of values.
    - Comparative description: articles where authors included a systematic comparison between software management tools by assigning values of some kind in the comparison.
    - Quantitative comparison: articles where authors included a systematic comparison by assigning numerical values to individual parameters in order to obtain assessments within a pre-established scale.

2. The second level corresponds to the functions or items that the authors used as indicators in the study. This level focused on considering which specific functions were used in each of the overall studies, and under which parameters.

Most of the articles (19, 51.4%) chose a comparative description approach to analysis. There is not a significant difference between the areas of Library and Information Science and Biomedicine (10 articles in the first case, 9 in the second). However, the concept of comparative description, or how it is approached in the publication, varies among the analyzed articles. Several authors conducted a comparative study with a narrative structure, either when describing each software tool or when describing a function or characteristic. A few articles focused exclusively on contrasting specific functions, such as database query capabilities (Gomis, Gall and Brahmi, 2008). And a quantitative methodology was used in articles focused on the mistakes found in generated bibliographies (see *infra*).

In any case, the key element of this type of study is the use of comparative function tables. Only 18 articles (48.6%, 3 of which were quantitative comparisons, *see infra*) offered tables, but almost always with heterogeneous contents and organization. The use of tables became widespread in the second half of the decade of 1990, but because these studies did not use indicators of quantitative assessment, their value is limited in establishing the presence or absence of a function. These tables of presence/absence of functions are a first step to the systematic assessment of software tools, but the reviewed articles did not go any further. Another lack shown by this kind of article is that authors did not define a target group of users; therefore, they fail to establish which functions or characteristics are the most wanted and as a result they end up replicated models based on the technical features already present in the reference management software. Cibarelli (1995) was the first to approach the question using a comprehensive survey sent to users, but the survey was focused on how users assess the reference software management tools rather than in developing a model and evaluation method to use with the collected data.

Thus, the characteristics, functions, and features offered by the reference management software tools were used to establish the assessment models used in the previously described articles. It is a model based on reviewing functions that are not ideal or based on user needs, but instead based on the existing features of the tools. The categories used and compared for analysis included data search, edition and capture, citation styles, word processor integration, data import and export, and interaction with social web services. Some studies evaluated the information on the software manufacturer and application data, as well as the required operating system. By adopting this model, these studies only confirm or deny the presence or absence of a function, without performing load tests to see if it is properly fulfilling its purpose. The only load tests documented are those corresponding to database queries and to the generation of proper reference results according to styles, which match those articles using a quantitative comparison model (*see infra*). The assessment criteria identified in the reviewed articles appear in table 2.

| General Criteria | Specific Criteria |
|---|---|
| **Search** | External databases and other sources<br>Academic engines<br>Own/house databases |
| **Data capture** | Capture references from external databases<br>Manual creation of references<br>Errors in imported files |
| **Reference and data edition** | Edit, modify and delete references<br>Duplicate detection<br>Capacity for global data changes<br>Annotation<br>Tagging |
| **Bibliography creation** | Creation of bibliographical list (text format)<br>Creation of HTML documents<br>Errors in applying bibliographic styles |
| **Authorities** | Creation and management of list of authorities |

| **Data import and export** | Import files coded in bibliographic exchange formats<br>Export files coded in bibliographic exchange formats |
|---|---|
| **File management** | Organization and management of PDF files<br>Metadata extraction of PDF files |
| **Bibliographic styles** | Number of available styles<br>Creation or import of new styles<br>Errors in applying styles |
| **Metadata** | Export/import to metadata schemes |
| **Word processor integration** | Integration of references in word processor documents<br>Automatic generation of bibliographies in word processors |
| **Diffusion and publishing** | Sending references and bibliographies by email<br>Sharing references in social networks<br>RSS. |
| **Collaborative work** | Syncronization with bibliographic web services<br>Creation and management of shared bibliographies<br>Creation of shared work groups<br>Social tagging<br>Work offline<br>Access control |
| **User interface** | Personalization<br>Ease of use<br>Usabilility<br>Records display options |
| **Reliability** | Reliability |
| **User help** | Technical and user documentation<br>Manuals<br>Tutorials |
| **Interaction with other software applications** | API<br>Web version<br>Mobile version |
| **Seller or provider** | Product cost<br>Supported operating systems<br>Interface translations<br>User attention and support<br>License type |

Table 2. Assessment criteria identified in the revised literature.

The assessment criteria identified correspond to an evaluation approach which stems from the ideal model of the desktop application, oriented towards personal work and formulated from the point of view of the expert in reference management. This has been the predominant focus in the field regarded as object of study: in 1988, eight attributes were established for "the perfect bibliographic software" (Brantz and Galla, 1988), whereas in 1993 a decalogue of questions was suggested to the user before they choose one or another tool (Jones, 1993). It was the common orientation until the beginning of the decade of 2010.

Regarding more specific criteria s, the most common aspects are those related to capture, edition and generation of bibliographic data. . In the 37 studies the authors usually focused on those same characteristics. The interest in reference annotation and PDF management capabilities began around 2010, after the function had been incorporated in reference management tools (*Zotero* appears in 2006). Bibliography making and word processor integration were two processes that became increasingly important at the end of the 1990s. Citation style management is another evaluation parameter that remained constant in the reviewed articles, albeit without assessing more specialized attributes such as the ability to interact with *Citation Style Language* (CSL) repositories.

The assessment of bibliographic generation capabilities and website integration made a late appearance in the articles reviewed. Not until 2002, a decade after the advent of the Web, did we find a research article (Koopman, 2002) that reviewed the quality and results of bibliographic generation. Only after the assessments incorporated for data publication and dissemination functions in social media and collaborative environments. *CiteULike* and *Connotea* were launched in 2004; *Bibsonomy* and *Zotero* in 2006. A year later, a first descriptive-comparative article was identified, including web applications with dissemination and collaborative work functions (Giménez and Tramullas, 2007). In 2006, Hendry, Jenkins, and McCarty suggested a theoretical model to develop collaborative bibliographies, considering various collaborative work and bibliographic development and maintenance scenarios. However, the assessment of these functions kept the previous focus of enumerating and affirming or denying them, without offering measurable related criteria and data.

Although there are some basic studies on user opinion (Lorenzetti and Ghali, 2013), evaluations of usability and ease of use were not found in the reviewed articles. This provides an interesting contrast to the high number of published articles on usability in OPACs and in library computerization systems.

The second most common model used by authors was a simple description, which enumerated the characteristics and features of the reference management software tools (13 articles, 35.1%). The descriptive depth of functions and characteristics was highly heterogeneous in these articles, which means that some of them were limited to an enumeration of functions without providing further verification or cross-checking. As happened with comparative descriptions, neither context of use nor a target group of users was defined. Thus, the only value of these articles is as an overview of the applications, functions, and market as they existed at a given point in time.

A quantitative comparison was the least common approach (5 articles, 13.5%). Only one of the articles reviewed offered a basic quantitative approach (Basak, 2014); the other four (Garfield, Flanagan and Fox, 1989; Kessler and Van Ullen, 2005; Gilmour and Cobus-Kuo, 2011; Homol, 2014) actually conducted a quantitative assessment regarding the number of mistakes managers made when generating bibliographic references, considering various citation styles. Thus, these would be partial comparisons, exclusively based on a specific aspect of reference managers.

The review of the applied criteria in the articles revealed two distinct stages in the evolution of reference management software:

1. A first stage, up until 2006, wherein tools were understood as desktop applications, for personal purposes, with a traditional approach regarding bibliographic capture, management and generation, and therefore directed toward traditional publishing.
2. A second stage, beginning in 2007, where the emergence of web 2.0. collaboration and information dissemination through a number of web technologies was progressively incorporated.

An important aspect yet to be mentioned on this article is the support provided by librarians to their users, and as a result librarians become well-acquainted with reference management software. Recently, a handful of articles have been published that address this topic. MacMinn (2011) conducted a library website review and identified 111 websites providing information about reference management software. Childress (2011) described how the library at *Penn State University* developed a working group to analyze patterns of reference manager utilization by its users. The analysis revealed that these tools are used within the general context of academic work including educating users about plagiarism, citation styles, and the criteria they can use to select reference management tools for their needs. More recent research by Salem and Fehrmann (2013) used focus groups to study how college students use reference management software.

**6. Conclusions**

This review of comparative studies on reference management software provides grounds to claim that there is not a common or standardized method of analysis. The models applied by the authors corresponded to approaches based on their assumed expert knowledge of the tools themselves. Obviously, this approach overlooks the possibility of building ideal theoretical manager models, based the needs and actual activities of the users. In addition, authors did not take into account the common rules and standards for software quality evaluation: they did not use inspection techniques or defined metrics, as established by the standards ISO/IEC 9126, 14598 and 25010. The current revised standard corresponds to ISO/IEC 25010 (2011), which includes the contents of the classic ISO/IEC 9126-1 (2001). A proper evaluation reference software management tools should include a quality model, characteristics, requirements, and metrics indicated in the standard. It can be stated that the reviewed articles do not show a rigorous methodology in their approach and execution of evaluations. In conclusion, the working hypothesis presented at the beginning of this article can be considered valid.

Another observation, inferred from the review, is that there is not a standard definition or clearly described concept of "reference management software" beyond the generic claims and classic definitions. Authors of the 37 articles seemed to accept tools themselves as an ideal model and did not devote any section of their evaluation to elaborate on this concept. It should be highlighted that tools have evolved as well as the technological context. Reference management software can no longer be defined in the traditional sense provided by the reviewed literature. Currently, it is an integrated tool of information management providing support to workflows of scientific research in any given area. Fourie suggested considering reference management software as a specific type of personal information manager, with the ability to combine the information they contain with organizations and visualizations such as *Topic Maps* (Fourie, 2011). Hull, Petiffer and Kell (2008) went further when reviewing the integration of digital libraries into web-based reference managers. The problems they identified to make good use of their digital content were only solved partially, through better and more powerful reference managers, and through information personalization and socialization.

Finally, Library Science has failed to lead in this area of research and has not offered significant contributions to it, neither in theoretical nor technical aspects. The reviewed literature shows a passive attitude, merely intended as a revision, of the work of third parties. More proactive approaches did not arise until 2010. These approaches observe users and their information behavior, to plan and carry out actions with and on reference managers. As pointed out by Kathleen and Kaye (2011: 208): "...we are freed to spend less energy teaching the specifics of citation styles and more time on not only why it is essential to properly cite but to introduce more advanced information management skills..."